\documentclass{article}
\usepackage{txfonts}
\usepackage{stmaryrd}
\usepackage{booktabs}
\usepackage{amssymb}
\usepackage{latexsym}
\usepackage{graphicx}

\makeatletter
    
    \newcommand{\Rmnum}[1]{\expandafter\@slowromancap\romannumeral #1@}
  \makeatother

\begin{document}
\begin{center}
\Large {Signal Photon Flux and Background Noise in a Coupling
Electromagnetic Detecting System for High Frequency Gravitational
Waves (revised version)}
\end{center}

\begin{center}
F.Y. Li$^{1,a}$ \mbox{}~~ N. Yang$^{1,b}$ \mbox{}~~ Z.Y.
Fang$^{1,c}$ \mbox{}~~R.M.L. Baker Jr.$^{2,d}$ \mbox{}~~ G.V.
Stephenson$^{3,e}$ \mbox{}~~ \\ H. Wen$^{1,f}$
\end{center}

$^{1}$Department of Physics, Chongqing University, Chongqing 400044,
P. R. China

\mbox{$^{2}$GRAWAVE{\textregistered} LLC, 8123 Tuscany Avenue, Playa
del Rey, California 90293, USA}

$^{3}$Seculine Consulting, P0 Box 925, Redondo Beach, CA 90277, USA

\begin{center}
\textbf{Abstract}
\end{center}

\large{A coupling system between Gaussian type-microwave photon
flux, static magnetic field and fractal membranes (or other
equivalent microwave lenses) can be used to detect high-frequency
gravitational waves (HFGWs) in the microwave band. We study the
signal photon flux, background photon flux and the requisite minimal
accumulation time of the signal in the coupling system. Unlike pure
inverse Gertsenshtein effect (G-effect) caused by the HFGWs  in the
GHz band, the the electromagnetic (EM) detecting scheme (EDS)
proposed by China and the US HFGW groups is based on the composite
effect of the synchro-resonance effect and the inverse G-effect. Key
parameters in the scheme include first-order perturbative photon
flux (PPF) and not the second-order PPF; the distinguishable signal
is the transverse first-order PPF and not the longitudinal PPF; the
photon flux focused by the fractal membranes or other equivalent
microwave lenses is not only the transverse first-order PPF but the
total transverse photon flux, and these photon fluxes have different
signal-to-noise ratios at the different receiving surfaces.
Theoretical analysis and numerical estimation show that the
requisite minimal accumulation time of the signal at the special
receiving surfaces and in the background noise fluctuation would be
$\sim10^3-10^5$ seconds for the typical laboratory condition and
parameters of $h_{r.m.s.}\sim10^{-26}-10^{-30}$ at 5GHz with
bandwidth $\sim$1Hz. In addition, we review the inverse G-effect in
the EM detection of the HFGWs, and it is shown that the EM detecting
scheme based only on the pure inverse G-effect in the laboratory
condition would not be useful to detect HFGWs in the microwave
band.

 \small{\noindent{$^a$cqufangyuli@hotmail.com,
$^b$cquyangnan@cqu.edu.cn $^c$zyf@cqu.edu.cn
$^d$DrRobertBaker@GravWave.com $^e$E-mail: seculine@gmail.com $^f$
wenhaowww@yahoo.com.cn}}

\newpage
\noindent\Large{ \Rmnum{1}. Introduction.}
 \large{\mbox{}}\\
 \mbox{}\\

The first mention of high-frequency gravitational waves (HFGWs)was
during a lecture in 1961 by Robert L. Forward[1]. The lecture was
based upon a paper concerning the dynamics of gravity and Forward's
work on the Weber Bar. The first actual publication concerning HFGWs
was in mid 1962 when M.E Gertsenshtein[2] authored the pioneering
paper entitled ``wave resonance of light and gravitational waves"
(it is often called Gertsenshtein effect). The next publication was
in August of 1964 when L.Halpern and B.Laurent[3]; they suggested at
some earlier stage of development of the universe (the big bang)
conditions were suitable to produce strong relic gravitational
radiation. They then discuss ``short wavelength" or HFGWs and even
suggest a `laser' generator of HFGWs analogous to a laser for EM
`generation'. In 1968 R.A.Isaason authored papers[4,5] concerned
with ``Gravitational Radiation in the Limit of High Frequency".
L.P.Grishchuk and M.V.Sazhin in the  periods of 1974-1975 disccussed
a scheme on ``Emission of gravitational waves by an electromagnetic
cavity and detection"[6,7], which also involved HFGWs. In 1974
G.F.Chapling, J.Nuckolls and L.L.Woods[8] suggested the generation
of HFGWs by nuclear explosions and in 1978. V.B.Braginsky and
V.N.Rudenko discussed detection and generation of the HFGWs [9]. In
1979 S.W.Hawking and W.Isreal[10] presented an actual definition for
HFGWs having frequencies in excess 100KHz. However, genuine
attention to HFGWs was occurred from the 1990's for the following
reasons:

(1)The maximal signal and peak of the relic GWs, expected by the
quintessential inflationary models (QIM) [11-15] and some string
cosmoogy scenarios[16-18], may be firmly localized in the GHz band,
and their root-mean-square (r.m.s) values of the dimensionless
amplitudes might reach up to $\sim10^{-30}-10^{-33}$. Such works
continue today.

(2) The thermal motion of plasma of stars, the interaction of the EM
waves with interstellar plasma and magnetic fields, and the evaporation
of primordial back holes[19], are possible means to generate the
HFGWs.

(3)Study of nano-piezoelectiric resonator scheme[20], high-energy
particle beam[21-25] and the construction of the LHC[26] are possible methods to
produce HFGWs. Their frequencies may reach up to $10^9$ Hz and
higher.

(4) Some HFGW detectors have already been constructed and more have been proposed. The
constructed HFGW detectors include a toroidal waveguide scheme[27,28]
and a coupled superconducting spherical cavities system[29,30].
Proposed detecting schemes include small
laser interferometers detectors[31] and the coupling system of Gaussian beam, static magnetic and fractal membranes [32]. In Table 1 we list some possible HFGW sources and their major mechanisms.\\

In this paper our attention is focused on signal photon flux,
the background photon flux(BPF) and their signal-to-noise ratios in
the coupling EM detection scheme. We compute the signal photon
fluxes, the signal-to-noise ratios, and discuss displaying condition
and the requisite minimal accumulation time of the signal in the
background noise fluctuation. In addition, we review the inverse
G-effect in the EM detection of the HFGWs. It is shown that the pure
inverse G-effect in the laboratory condition cannot by itself detect the
expected HFGWs, but the current EM detecting scheme might greatly
improve detecting sensitivity and narrow the gap between the
theoretical estimation  of the expected HFGWs and  the possibility of their detection.\\

The outline of this paper is the following; In Sec. \Rmnum{1} we
present a brief history of the HFGWs research, including analyses of
some possible HFGW sources. In sec. II we review the detecting
scheme based on the pure inverse G-effect. In Sec.III we discuss the
EM perturbation generated by the HFGW in coupling system between the
static magnetic and the plane EM wave. In sec.IV we study the EM
perturbative effect of the HFGW in the coupling system between the
Gaussian type-microwave photon flux, the static magnetic field, and
the fractal membranes (or other equivalent microwave lenses), and
give theoretical analysis, numerical estimations and a brief review
to the role of the fractal membranes or other equivalent microwave
lenses. Our conclusions are summarized in Sec
     V.

\mbox{}\\
\mbox{}\\
\noindent\Large{\Rmnum{2}. Detecting scheme based on the inverse
Gertsenshtein

effect.}
 \large{\mbox{}}\\

\begin{table}[htbp]
\caption{Some possible HFGW sources and relevant paramenters}
\begin{center}
\begin{tabular}{p{60pt}p{150pt}p{100pt}p{100pt}}

\hline Sources & Amplitude & Frequency &Characteristic\\

\hline HFRGWs in the quintessential inflationary models [11-15] &$h_{rms}\sim10^{-30}-10^{-32}$ &$\nu\sim10^{9}-10^{10}$Hz&Stochastic background \\

\hline HFRGWs in some string cosmology scenarios [16-18]&
$h_{rms}\sim10^{-30}-10^{-34}$ &$\nu\sim10^{8}-10^{11}$Hz& Stochastic background\\

\hline Solar plasma \mbox{}[19]&
$h_{rms}\sim10^{-39}$& $\nu\sim10^{15}Hz$ & On the earth \\

\hline High-energy  particles (e.g., Fermi ring) [24]&
$h_{rms}\sim10^{-39}-10^{-41}$& $\nu\sim10^{4}Hz-10^5Hz$& On the
center, the frequency depends on the rotating frequency of the
particles in the Fermi
ring  \\

\hline Stanford Linear Collider (SCL)[21]& $h_{rms}\sim10^{-39}$&
$\nu\sim10^{23}Hz$& On the collision center, the frequency depends
on the
self-energy and the Lorenty factor of high-energy $e^+ e^-$ beams \\

\hline The Large Hadron Collider (LHC) [26]&
  & & This is a continuous spectra of high-frequency gravitons, only integrals for the total spectra distribution range might provide an indirect effect.\\

\hline Nano-piezoelectric crystal array (size of $\sim$ 100m)[20]&
$h_{rms}\sim10^{-28}-10^{-31}$& $\nu\sim10^{9}-10^{10}Hz$& On the
wave zone, effective cross section of the gravitational
radiation would be less than 0.01$m^2$\\

\hline

\end{tabular}
\label{tab1}
\end{center}
\end{table}
It is well known that if an electromagnetic wave (EMW) propagates in
a transverse homogeneous static magnetic field, it can generate the
gravitational wave (GW). This is just the G-effect[2]. Then
converting probability of the EMW (photons) into the GW (graviton)
is given by [33,34] (in CGS units)
\begin{equation}
\label{eq1}
P\approx 4\pi GB^2L^2/c^4,
\end{equation}
where $G$ is Newton's gravitational constant, $B$ is the static
magnetic field. Contrarily, if a GW passes through a transverse
homogeneous static magnetic field, then it can generate an EMW
(photon flux), which propagates only in the same and in the opposite
propagating directions of the GW. The latter is weaker than the
former or is absent. This is just the pure inverse G-effect [33,35].
Whether the G-effect or its inverse effect, the conversion rate
between the GWs (gravitons) and the EMWs (photons) is extremely low.
For example, if $B=10T=10^{5}$Gauss, $L=10m=1000$cm, from Eq.(1), we
have
\begin{equation}
\label{eq2}
P\approx 1.0\times 10^{-32}\mbox{.}
\end{equation}
For the EM perturbative effect caused by the GWs in the EM fields,
one's attention is often focused to the inverse G-effect. In order
to consider the pure inverse G-effect in the laboratory size, the
wavelength of GWs should be the comparable with the laboratory
dimension. Thus the high-frequency GWs (HFGWs) in the microwave band
($\sim $10$^{8}$-10$^{10}$Hz) would be suitable researching object.
In fact, physical foundation of the G-effect is the Einstein-Maxwell
equations in the weak field condition, while the physical foundation
of the inverse G-effect is classical electrodynamics in curved
spacetime. If a circular polarized HFGW passes through the
transverse homogenous static magnetic field, according to the
electrodynamical equations in curved spacetime, the EMW produced by
the interaction of the HFGW with the static magnetic field can be
given by [32,35] (in order to compare possible experimental effect,
from now, we use MKS units).
\begin{equation}
\label{eq3}
\vec {E}^{(1)}\approx A\hat {B}_y^{(0)} k_g cz\exp [i(k_g z-\omega _g t)],
\end{equation}
\begin{equation}
\label{eq4}
\vec {B}^{(1)}\approx A\hat {B}_y^{(0)} k_g z\exp [i(k_g z-\omega _g t)],
\end{equation}
where $\vec {E}^{(1)}$ and $\vec {B}^{(1)}$ are parallel to the
xy-plane and $\vec {E}^{(1)}\bot \vec {B}^{(1)}$. We also assume
$A=A_\oplus =A_\otimes =\left| {h_\oplus } \right|=\left| {h_\otimes
} \right|,$ as the amplitudes of the HFGW with two
polarization states, and the superscript (0) denotes the background
EM fields, the notation \^{} indicates the static EM fields,
respectively. Here we neglected the EMW propagating along the
negative direction of the z-axis, because it is often much less than
the EMW propagating along the positive direction of the z-axis.
Eqs.(3) and (4) show that such perturbative EM fields have a space
accumulation effect ($\propto z)$ in the interacting region: this is
because the GWs (gravitons) and EMWs (photons) have the same
propagating velocity in a vacuum, so that the two waves can generate an optimum
coherent effect in the propagating direction [33,35]. From Eqs. (3)
and (4), the power flux density of the EMW in the terminal receiving
surface (z=L) will have maximum (z=L, see Figure 1)
\begin{equation}
\label{eq5} {u_{em}} = 1/{\mu _0} \cdot |{\vec E^{(1)}} \times {\vec
B^{(1)}}| \approx 1/{\mu _0} \cdot {(A\hat B_y^{(0)}{k_g}L)^2}c.
\end{equation}

\begin{figure}[htbp]
\centerline{\includegraphics[scale=0.5]{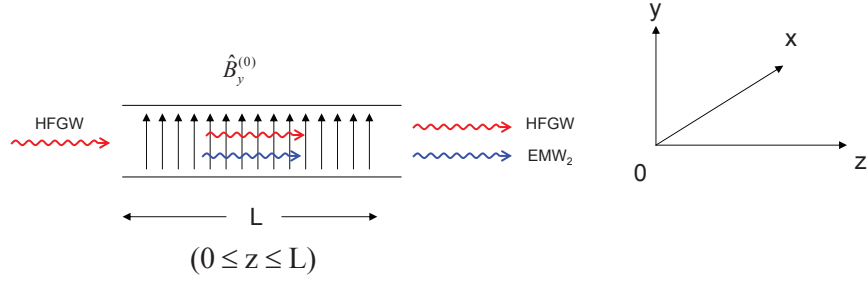}} \label{fig1}
\caption{ If a HFGW passes through a static magnetic field
$\hat{\vec{B}}_y^{(0)}$, the interaction of the HFGW with the static
magnetic field will produce an EMW, where L is the interacting
dimension between the HFGW and the static magnetic field. The
EMW$_{2}$ has maximum in the terminal position (Z=L) of the
interacting volume due to the space accumulation effect in the
propagating direction (the z-direction).}
\end{figure}

In order to compare and analyze the EM perturbative effect under
typical laboratory conditions, we choice following typical
parameters,
\begin{equation}
\label{eq6}
\begin{array}{l}
 \hat B_y^{(0)} = 10{\mathop{\rm T}\nolimits} ,{\rm{    L}} = {\rm{1}}0{\rm{m,    }} \\
  \\
 {\nu _e} = {\nu _g} = 5{\rm{GHz (}}{\lambda _g} = 0.06m,{k_e} = {k_g} = \frac{{2\pi }}{\lambda } \approx 100{\rm{),}} \\
  \\
 h\nu  = 3.3 \times {10^{ - 24}}J({\rm{energy~ of~ single~ photon}}),{\rm{ }} \\
  \\
 A \approx {h_{rms}} = \hat h = 10^{-26}~to~10^{-30}, \\
  \\
 \Delta s = 0.1 \times 0.1 = 0.01{{\rm{m}}^2}{\rm{ (typical~ receiving~ surface),}} \\
 \end{array}\
\end{equation}
where $\Delta s$ is also the cross section of the interacting
region. If $\hat{h}=h_{rms}=10^{-30}$, then the total power flux
passing through $\Delta s$ in the terminal position (z=L) is
\begin{equation}
\label{eq7} U_{em}^{(2)} = {u_{em}}\Delta s = \frac{{\rm{1}}}{{{\mu
_0}}}{(A\hat B_y^{(0)}{k_g}L)^2}c\Delta s \approx 2.3 \times {10^{ -
40}}W,
\end{equation}
where the superscript (2) denotes the second-order perturbative EM power
flux. Therefore, corresponding second-order perturbative photon flux (in
quantum language) will be
\begin{equation}
\label{eq8} N_\gamma ^{(2)} = U_{em}^{(2)}/\hbar {\omega _e} \approx
2.3 \times {10^{ - 40}}/3.3 \times {10^{ - 24}} \approx 7.0 \times
{10^{ - 17}}{s^{ - 1}}.
\end{equation}
For an HFGW of ${\nu _g} = 5GHz, \hat h = {10^{ - 30}}$, the total
power flux passing through the $\Delta s$ is given by [36]
\begin{equation}
\label{eq9} {U_{gw}} = {u_{gw}}\Delta s = \frac{{{c^3}}}{{8\pi
G}}{\omega ^2}{A^2}\Delta s \approx 1.6 \times {10^{ - 7}}W,
\end{equation}
Thus corresponding graviton flux would be
\begin{equation}
\label{eq10} {N_g} = {U_{gw}}/\hbar \omega  \approx 4.8 \times
{10^{16}}{s^{ - 1}}.
\end{equation}
Because the power fluxes, Eq.(\ref{eq7}) (including the photon flux, Eq.(\ref{eq8})) is
proportional to the amplitude squared of the HFGW, the second-order
perturbative photon flux (PPF) exhibits a very small value.

From Eqs.(\ref{eq7})-(\ref{eq10}), we obtain the conversion rate of the HFGW (gravitons)
into the EMW (photons) as follows
\begin{equation}
\label{eq11} P \approx {U_{em}}/{U_{gw}} = {N_\gamma }/{N_g}=
\frac{{2.3 \times {{10}^{ - 40}}}}{{1.6 \times {{10}^{ - 7}}}} \\
= \frac{{7 \times {{10}^{ - 17}}}}{{4.8 \times {{10}^{16}}}} \approx
1.4 \times {10^{ - 33}}.
\end{equation}
Eqs.(\ref{eq2}) and (\ref{eq11}) show that the conversion rates of the EMW (photons)
into the HFGW (gravitons) and the contrary process have similar orders of
magnitude. Thus, in order to obtain a second-order perturbative photon, from
Eq. (\ref{eq8}), the signal accumulation time would be, at least
\begin{equation}
\label{eq12} \Delta t \approx 1/N_r^{(2)} \approx \frac{1}{{7 \times
{{10}^{ - 17}}}} \approx 1.4 \times {10^{16}}{\rm{s}}{\rm{.}}
\end{equation}
This is a very huge time interval. Eqs.(\ref{eq11}) and (\ref{eq12})
also show that the conversion rate of the HFGW (gravitons) into the
EMW (photons) is extremely low. Thus the PPF in the pure inverse
G-effect cannot cause a detectable signal or observable effect in
the laboratory condition. Nevertheless, for some astrophysical and
cosmological processes, it is possible to cause interesting
phenomena, because the very large EM fields (including plasma) and
very strong GWs (including low frequency GWs) often occur
simultaneously and these fields extend over a very large area
[15,37,38].

From Eqs. (\ref{eq5}) (\ref{eq7}),(\ref{eq8}) and (\ref{eq12}), one
finds,\mbox{}\\
\begin{equation}
\label{eq13} \begin{array}{l}
 {\rm{if~    }}\hat h = {10^{ - 26}},{\rm{~  then   ~  }}N_\gamma ^{(2)} \approx 7 \times {10^{ - 9}}{{\rm{s}}^{ - 1}}{\rm{   ~ and ~ }}\Delta t \approx 1.4 \times {10^8}{\rm{s,}} \mbox{}\\\mbox{}\\
 {\rm{      }}\hat h = {10^{ - 24}},{\rm{  ~  then     ~}}N_\gamma ^{(2)} \approx 7 \times {10^{ - 5}}{{\rm{s}}^{ - 1}}{\rm{    ~and ~}}\Delta t \approx 1.4 \times {10^4}{\rm{s}}{\rm{.}} \\
 \end{array}
\end{equation}\mbox{}\\\mbox{}\\
Such results show that even if $\hat {h}=10^{-24}$, it is still
difficult to detect the HFGWs by the inverse G-effect in the
laboratory condition. In other words, in order to generate an
observable effect in such EM system, the amplitude of the HFGW of
$\nu _g =5GHz$ must be larger than $\hat {h}=10^{-24}$ at least.
Unfortunately, so far as, we know there are no those HFGWs as strong
as $\hat {h}=10^{-24}$ or larger, though the EM system based on the
pure inverse G-effect in the high-vacuum and ultra-low-temperature
condition has a very good low
noise environment. Therefore the EM detecting scheme based the pure inverse G-effect in the laboratory condition would not be available to detect HFGWs in the microwave band.\\

\noindent\Large{\Rmnum{3}. The perturbative photon fluxes in
coupling system between the static magnetic field and the
\mbox{}plane EMW. }
 \large{\mbox{}}\\

The classical and semi-classical description and linear quantum theory
all showed [33,39] that the interaction cross section between the GW
(gravitons) and the EMW(photons) in a strong background static
magnetic field (virtual photons) will be much larger than that in
the pure inverse G-effect. In other words, the strong background
static magnetic field provides a catalyst to greatly enhance the
resonant effect between the EMW (the photons) and the GW
(gravitons). However, the presence of background EMW (the background
photon flux) will generate a large photon flux noise. If the
perturbative photon flux (PPF, i.e., signal photon flux) and the
background photon flux (BPF) have the same or the very similar
physical behaviors (e.g., propagating direction, distribution, decay
rate, etc.), then the PPF will be swamped by the BPF. The coupling
system between a plane EMW and the static magnetic field is just
this case (see Fig.2), which will have the same or very similar sensitivity  as the inverse G-effect. We
assume the power of the background EMW is 10W, and it is limited in
the cross section of $\Delta s=0.1\times 0.1=0.01\mbox{m}^2$.
Because the power flux of the plane EMW is distributed homogeneously
in the cross
section $\Delta s$, then\mbox{}\\
\begin{equation}
\label{eq14}
\begin{array}{l}
 \left\langle {{P_{em}}} \right\rangle  = {\mathop{\rm Re}\nolimits} \left( {\frac{1}{{2{\mu _0}}}E_x^{*(0)}B_y^{(0)}} \right)\Delta s = \frac{1}{{2{\mu _0}}}\frac{{E_x^{{{(0)}^2}}}}{c}\Delta s = 10{\rm{W,}} \\
  \\
 {\rm{and    ~~~~~~~~~~~~~~   |}}\vec E_x^{(0)}| \approx 8.7 \times {10^2}V{m^{ - 1}}. \\
 \end{array}
\end{equation}

\begin{figure}[htbp]
\centerline{\includegraphics[scale=0.45]{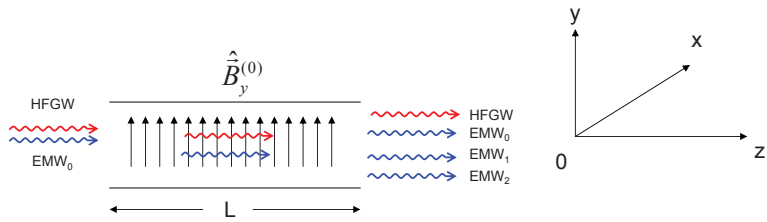}} \label{fig2}
\caption{ If the HFGW and the $EMW_0$ pass simultaneously through
the transverse static magnetic field  , under the resonant state
($\omega_e=\omega_g$), the first-order perturbative EMW ($EMW_1$,
i.e., ``the interference term") and the second-order perturbative
EMW (the $EMW_2$) can be generated. However, because the $EMW_1$ and
the $EMW_0$ have the same propagating direction and distribution,
and $EMW_1$ is often much less than the $EMW_0$, the $EMW_1$ will be
swamped by the $EMW_0$}
\end{figure}

Total background photon flux passing through the cross section
$\Delta s$ will be
\begin{equation}
\label{eq15} N_\gamma ^{(0)} = 10/\hbar {\omega _e} =
\frac{{10}}{{3.3 \times {{10}^{ - 24}}}} \approx 3.0 \times
{10^{24}}{{\rm{s}}^{{\rm{ - 1}}}}.
\end{equation}
Then corresponding first-order perturbative power flux in the
z-direction is

\begin{equation}
\label{eq151} U_z^{(1)} = \frac{1}{{2{\mu _0}}}{[({{\vec E}^{(1)}}
\times \vec B_y^{(0)}) + (\vec E_x^{(0)} \times {{\vec
B}^{(1)}})]_{{\omega _e} = {\omega _g}}}\Delta s .
\end{equation}
\[
\begin{array}{l}

  = {\mathop{\rm Re}\nolimits} [\frac{1}{{{\mu _0}}}{E^{(1)*}}B_y^{(0)}]\cos \beta \cos \delta  \cdot \Delta s \\
  = {\mathop{\rm Re}\nolimits} [\frac{1}{{{\mu _0}c}}{E^{(1)*}}E_x^{(0)}]\cos \beta \cos \delta  \cdot \Delta s \\
  = \frac{1}{{{\mu _0}c}}|{{\vec E}^{(1)}}||\vec E_x^{(0)}|\cos \beta \cos \delta  \cdot \Delta s, \\
 \end{array}
\]
where $\delta $ is the phase difference between the HFGW and the
background EMW$_{0}$, $\beta $ is the angle between $\vec
{E}^{(1)}\mbox{ and }\vec {E}_x^{(0)} $ or $\vec {B}^{(1)}\mbox{ and
}\vec {B}_y^{(0)} $ (see Fig.3). Here $\delta =0$ and $\beta =0$
will always be possible by regulating the phase and the polarization
directions of the background EMW$_{0}$. Then
the HFGW and the EMW will have the best matching state, i.e., \mbox{}\\
\begin{equation}
\label{eq16} {\left. {U_z^{(1)}} \right|_{\scriptstyle \delta  = 0
\hfill \atop
  \scriptstyle z = L \hfill}} = U_{z{\rm{ max}}}^{(1)} = {\mathop{\rm Re}\nolimits} [\frac{1}{{{\mu _0}c}}{E^{(1)*}}E_x^{(0)}] \Delta s\approx 6.9 \times {10^{ - 20}}{\rm{W}}{\rm{.}}
\end{equation}
\mbox{}\\
Then the corresponding first-order PPF will be\mbox{}\\
\begin{equation}
\label{eq17} N_z^{(1)} = U_z^{(1)}/\hbar {\omega _e} \approx 6.9
\times {10^{ - 20}}/3.3 \times {10^{ - 24}} \approx 2.1 \times
{10^4}{{\rm{s}}^{{\rm{ - 1}}}}.
\end{equation}
\mbox{}\\
Thus the total photon flux passing through $\Delta s$ is about\mbox{}\\
\begin{equation}
\label{eq18} {N_z} = N_z^{(0)} + N_z^{(1)} + N_z^{(2)} \approx (3.0
\times {10^{24}} + 2.1 \times {10^4} + 7.0 \times {10^{ -
17}}){{\rm{s}}^{{\rm{ - 1}}}}.
\end{equation}
\mbox{}\\
In this case the ratio of $N_z^{(1)} $and $N_z^{(0)} $is roughly\mbox{}\\
\begin{equation}
\label{eq19} {\sigma _1} = N_z^{(1)}/N_z^{(0)} \approx \frac{{2.1
\times {{10}^4}}}{{3.0 \times {{10}^{24}}}} \approx 7.0 \times {10^{
- 21}},
\end{equation}
\mbox{}\\
This is also very small value, and at the same time,\mbox{}\\
\begin{equation}
\label{eq20} {\rm{   }}{\sigma _2} = N_z^{(2)}/N_z^{(1)} \approx
\frac{{7.0 \times {{10}^{ - 17}}}}{{2.1 \times {{10}^4}}} \approx
3.3 \times {10^{ - 21}},
\end{equation}
\mbox{}\\
i.e., the second-order PPF is much less than the first-order PPF,
while the first-order PPF is much less than the background photon
flux (BPF). This means that if an EM detecting system contains
simultaneously the static magnetic field and the EMW, then the
interaction cross section between the GW (gravitons) and the EMW
(photons) will be much larger than that in the pure inverse
G-effect. The classical description and linear quantum theory for
such a property have good self-consistency [33,39].

However, Eqs. (3),(4),(14),(16),(18) show that the first-order PPF
(signal) and the BPF (noise) have the same propagating direction and
distribution, and the BPF is much larger than the PPF, so that the
PPF will be swamped by the BPF. In this case the PPF has no direct
observable effect. According to Eqs. (3), (4), (\ref{eq16}) and
(\ref{eq17}), one finds
\begin{equation}
\label{eq21}
\begin{array}{l}
 {\rm{if~  }}\hat h = {10^{ - 26}},{\rm{    then  ~  }}N_z^{(1)} \approx 2.1 \times {10^8}{{\rm{s}}^{ - 1}},{\rm{ }} \\
 {\rm{if ~ }}\hat h = {10^{ - 25}},{\rm{    then   ~ }}N_z^{(1)} \approx 2.1 \times {10^9}{{\rm{s}}^{ - 1}}. \\
 \end{array}
\end{equation}
For example, if $\hat {h}=10^{-26}$, in order to displaying
first-order PPF, $N_z^{(1)} \Delta t$ must be effectively larger
than the background noise fluctuation $\sqrt {N_z^{(0)} \Delta t} $,
i.e.,

\begin{equation}
\label{eq22}
\begin{array}{l}
 N_z^{(1)}{(\Delta t)^{\frac{1}{2}}} > \sqrt {N_z^{(0)}} , \\
  \\
 {\rm{then ~  }}\Delta t > 6.8 \times {10^7}{\rm{s}}{\rm{,}} \\
 \end{array}
\end{equation}
where $N_z^{(0)}\Delta t$ is the expectation value with a Poisson
distribution of width $\sqrt{N_z^{(0)}}$. Eqs. (13) and (23) show
that such two schemes have similar detecting sensitivity. Thus,
detecting the HFGW of $\hat {h}=10^{-26}$ and $\nu =5GHz\mbox{
}$by such coupling EM system will also be very difficult.\\

\noindent\Large{\Rmnum{4}. Coupling system of the static magnetic
field and the

Gaussian type microwave photon flux}
 \large{\mbox{}}\\

The above discussion shows that in order to detect the first-order
PPF, one must find a special EM resonant system in which the PPF and
the BPF have very different physical behaviors, even if such
difference are only distributed in a few local regions.

Before we discuss the resonance effect of the HFGWs in the proposal
EM system, we give a general analysis of the photon flux. Here,
$\vec {E}^{(0)},\vec {B}^{(0)}$denote the background EM fields,
$\vec {E}^{(1)},\vec {B}^{(1)}$ the perturbative EM fields produced
by the interaction of the HFGW with the static magnetic field. Then
total EM power flux density is\mbox{}\\
\begin{equation}
\label{eq23}
\begin{array}{l}
 {{\vec u}_{em}} = \frac{1}{{{\mu _0}}}\mathop E\limits^ \to   \times \mathop B\limits^ \to   = \frac{1}{{{\mu _0}}}({{\vec E}^{(0)}} + {{\vec E}^{(1)}}) \times ({{\vec B}^{(0)}} + {{\vec B}^{(1)}}) \\
 {\rm{    }} = \frac{1}{{{\mu _0}}}{{\vec E}^{(0)}} \times {{\vec B}^{(0)}} + \frac{1}{{{\mu _0}}}({{\vec E}^{(0)}} \times {{\vec B}^{(1)}} + {{\vec E}^{(1)}} \times {{\vec B}^{(0)}}) + \frac{1}{{{\mu _0}}}{{\vec E}^{(1)}} \times {{\vec B}^{(1)}}. \\
 \end{array}
\end{equation}
\mbox{}\\
Thus, the corresponding total photon flux density will be\mbox{}\\
\begin{equation}
\label{eq24}
\begin{array}{l}
 {{\vec n}_\gamma } = \frac{1}{{\hbar {\omega _e}}}{{\vec u}_{em}} \\
 {\rm{  }} = \frac{1}{{{\mu _0}\hbar {\omega _e}}}({{\vec E}^{(0)}} \times {{\vec B}^{(0)}}) + \frac{1}{{{\mu _0}\hbar {\omega _e}}}({{\vec E}^{(0)}} \times {{\vec B}^{(1)}} + {{\vec E}^{(1)}} \times {{\vec B}^{(0)}}) \\
 {\rm{     }} + \frac{1}{{{\mu _0}\hbar {\omega _e}}}({{\vec E}^{(1)}} \times {{\vec B}^{(1)}}) \\
  \\
 {\rm{  }} = {{\vec n}^{(0)}} + {{\vec n}^{({\rm{1}})}} + {{\vec n}^{({\rm{2}})}} \\
 \end{array}
\end{equation}
\mbox{}\\
where
\begin{equation}
\label{eq25}
\begin{array}{l}
 {{\vec n}^{(0)}} = \frac{1}{{{\mu _0}\hbar {\omega _e}}}({{\vec E}^{(0)}} \times {{\vec B}^{(0)}}), \\
 {{\vec n}^{({\rm{1}})}} = \frac{1}{{{\mu _0}\hbar {\omega _e}}}({{\vec E}^{(0)}} \times {{\vec B}^{(1)}} + {{\vec E}^{(1)}} \times {{\vec B}^{(0)}}), \\
 {{\vec n}^{({\rm{2}})}} = \frac{1}{{{\mu _0}\hbar {\omega _e}}}({{\vec E}^{(1)}} \times {{\vec B}^{(1)}}). \\
 \end{array}
\end{equation}
Eq.(\ref{eq24}) and (\ref{eq25}) would be most general form of the
PPF and the BPF, where $\vec {n}^{(0)}$,$\vec {n}^{(1)}$and $\vec
{n}^{(2)}$express the BPF, the first-order PPF and the second-order
PPF densities, respectively. Since non-vanishing $\left| {\vec
{E}^{(0)}} \right|,\left| {\vec {B}^{(0)}} \right|$ are often much
larger than $\left| {\vec {E}^{(1)}} \right|,\left| {\vec {B}^{(1)}}
\right|$, we have\mbox{}\\
\begin{equation}
\label{eq26}
|{\vec n^{(0)}}| \gg |{\vec n^{(1)}}| \gg |{\vec n^{(2)}}|{\rm{.}}
\end{equation}
\mbox{}\\
4-1.In the case of the plane Electromagnetic Wave or the plane EMW.\mbox{}\\

\begin{figure}[htbp]
\centerline{\includegraphics[scale=0.5]{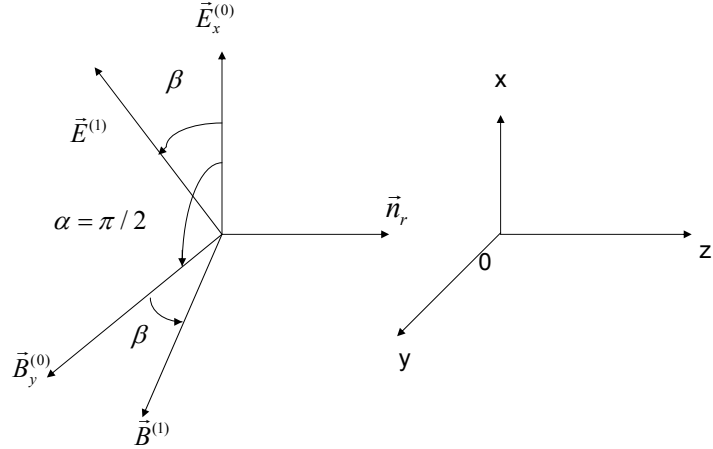}} \label{fig3}
\caption{ In the coupling system of the static magnetic field and
the plane EMW, $\left| {\vec {E}_x^{(0)} } \right|\mbox{ and }\left|
{\vec {B}_y^{(0)} } \right|${\small denote the background EM fields,
}$\left| {\vec {E}^{(1)}} \right|\mbox{ and }\left| {\vec {B}^{(1)}}
\right|${\small express the perturbative EM fields generated by the
direct interaction of the HFGW with the static magnetic field,
}$\vec {n}_\gamma ${\small is the total photon flux density.}}
\end{figure}

If the HFGW and the plane EMW$_{0}$ all propagate along the
z-direction, then Eq.(25) is deduced to (see Fig.3)\mbox{}\\
\begin{equation}
\label{eq27}
\begin{array}{l}
 {n_\gamma } = {\left\langle {{{\vec n}_\gamma }} \right\rangle _{{\omega _e} = {\omega _g}}} = \frac{1}{{2{\mu _0}\hbar {\omega _e}}}{\left\langle {(\vec E_x^{(0)} + {{\vec E}^{(1)}}) \times (\vec B_y^{(0)} + {{\vec B}^{(1)}})} \right\rangle _{{\omega _e} = {\omega _g}}} \\
 {\rm{  ~~~ }} = \frac{1}{{2{\mu _0}\hbar {\omega _e}}}\left\{ {\left| {\vec E_x^{(0)}} \right|\left| {\vec B_y^{(0)}} \right| + \left[ {|\vec E_x^{(0)}||{{\vec B}^{(1)}}|\sin \left( {\frac{\pi }{2} + \beta } \right)} \right.} \right. \\
 {\rm{ ~~ ~}} \\
 {\rm{   ~~~  }}\left. { + \left. {|{{\vec E}^{(1)}}||\vec B_y^{(0)}|\sin \left( {\frac{\pi }{2} - \beta } \right)} \right]\cos \delta  + |{{\vec E}^{(1)}}||{{\vec B}^{(1)}}|}
 \right\}.
 \\
 \end{array}
\end{equation}
\mbox{}\\
where the angular bracket denotes the average over time. For the
plane EMW in empty space, $B_y^{(0)} =E_x^{(0)} /c$,
$B^{(1)}=E^{(1)}/c$(in MKS units), then Eq.(\ref{eq27}) becomes
\begin{equation}
\label{eq28}
\begin{array}{l}
 {n_\gamma } = \frac{1}{{2{\mu _0}c\hbar {\omega _e}}}\left\{ {{{\left| {\vec E_x^{(0)}} \right|}^2} + 2|\vec E_x^{(0)}||{{\vec E}^{(1)}}|\cos \beta \cos \delta  + |{{\vec E}^{(1)}}{|^2}} \right\} \\
 {\rm{     ~~~    ~           }} = \frac{1}{{2{\mu _0}c\hbar {\omega _e}}}\left\{ {{{\left| {\vec E_x^{(0)}} \right|}^2} + 2\vec E_x^{(0)} \cdot {{\vec E}^{(1)}}\cos \delta  + |{{\vec E}^{(1)}}{|^2}} \right\} \\
  \\
 {\rm{        ~~~~            }} = {n^{(0)}} + {n^{(1)}} + {n^{(2)}}, \\
 \end{array}
\end{equation}
\mbox{}\\
where\mbox{}\\
\begin{equation}
\label{eq29}
\begin{array}{l}
 {n^{(0)}} = \frac{1}{{2{\mu _0}c\hbar {\omega _e}}}{\left| {\vec E_x^{(0)}} \right|^2}, \\
 {n^{(1)}} = \frac{1}{{{\mu _0}c\hbar {\omega _e}}}\vec E_x^{(0)} \cdot {{\vec E}^{(1)}}\cos \delta , \\
 {n^{(2)}} = \frac{1}{{2{\mu _0}c\hbar {\omega _e}}}|{{\vec E}^{(1)}}{|^2}. \\
 \end{array}
\end{equation}
\mbox{}\\
In fact, Eq.(\ref{eq29}) can also be expressed as
\begin{equation}
\label{eq30}
\begin{array}{l}
 {n^{(0)}} = \frac{1}{{2{\mu _0}c\hbar {\omega _e}}}{\left| {\vec E_x^{(0)}} \right|^2} = {{\dot N}_0}{\rm{(the~ background~ photon~ flux~ density)}} \\
 {n^{(2)}} = \frac{1}{{2{\mu _0}c\hbar {\omega _e}}}|{{\vec E}^{(1)}}{|^2} = {{\dot N}_{GW}}{\rm{(the~ second-order~ PPF ~density)}} \\
 \end{array}
 \end{equation}
 \mbox{}\\
while\mbox{}\\
\begin{equation}
\label{eq31}
\begin{array}{l}
 {n^{(1)}} = \frac{1}{{{\mu _0}c\hbar {\omega _e}}}\vec E_x^{(0)} \cdot {{\vec E}^{(1)}}\cos \delta  = 2{({{\dot N}_0}{{\dot N}_{GW}})^{\frac{1}{2}}}\cos \delta  = {{\dot N}_1} \\
 {\rm{        (the~ interference~ term,~ i}}{\rm{.e}}{\rm{., the~ first-order~ PPF~ density)}}{\rm{.}} \\
 \end{array}
\end{equation}
\mbox{}\\
Then, Eq.(29) can be re-written as\mbox{}\\
\begin{equation}
\label{eq32}
{n_\gamma } = {\dot N_0} + 2{({\dot N_0}{\dot
N_{GW}})^{\frac{1}{2}}}\cos \delta  + {\dot N_{GW}}.
\end{equation}
\mbox{}\\
After a long time interval $\Delta t$ the collected number of
photons at the detector or at the receiving surface would be\mbox{}\\
\begin{equation}
\label{eq33}
{N_d} = {n_\gamma }\Delta t = {\dot N_0}\Delta t + 2{({\dot
N_0}{\dot N_{GW}})^{\frac{1}{2}}}\cos \delta  \cdot \Delta t + {\dot
N_{GW}}\Delta t.
\end{equation}\mbox{}\\
Clearly, in the plane EMW case, the BPF, the first-order PPF and the
second-order PPF all propagate along the same direction, thus in any
region and at any receiving
surface\mbox{}\\
\begin{equation}
\label{eq34}
{\dot N_0} \gg 2{({\dot N_0}{\dot N_{GW}})^{\frac{1}{2}}} \gg {\dot
N_{GW}}
\end{equation}
\mbox{}\\
is always valid. In this case, it is very difficult to display the
first-order PPF effect (${n^{(1)}} = 2{({\dot N_0}{\dot
N_{GW}})^{\frac{1}{2}}}\cos \delta  = {\dot N_1}$) in an acceptable
signal accumulation time interval with the predicted total photon flux
background.

In the coupling system between the Gaussian type-microwave photon
flux (Gaussian beam (GB) is just one typical form of the Gaussian
type-microwave photon fluxes) and the static magnetic field, the
general expressions, Eqs.(25) and (26) are still valid. However,
they will be expressed as the different exact forms in the different
directions and the receiving surfaces, and the relative relation
between $n^{(0)}$ and $n^{(1)}$ would be different in the different
receiving surfaces, even then they can reach up a comparable order
of magnitude. This is worth consideration. The scheme from [32]
would be a useful candidate(see Fig. 4). Thus key parameters in the
scheme are the BPF and the first-order PPF in the special directions
and not the photon number. The former are vectors and have high
directivity. They decide the strength of the photon fluxes reaching
the detector or the receiving surface, position and bearings of the
detectors and
the signal-to-noise ratio (SNR) in the receiving surfaces.\mbox{}\\

\noindent4.2 Coupling system of the Gaussian-type microwave photon flux and the static magnetic field.\mbox{}\\

Unlike plane EMW, the GB has not only longitudinal BPF (the BPF in
the z-direction, i.e., the direction of its symmetrical axis) but
also the transverse BPF, although the latter is often less than the
former. The BPF in the transverse directions (e.g., the x-
and y- direction) decays as fast as the typical Gaussian decay rate.
Thus in the some special regions and directions, the effect of both
the PPF and the BPF would have a comparable order of magnitude.

For the GB with the double transverse polarized electric
modes[32,40] it
has\mbox{}\\
\begin{equation}
\label{eq35}
\begin{array}{l}
 {{\vec E}^{(0)}} = \vec E_x^{(0)} + \vec E_y^{(0)}, \\
 {{\vec B}^{(0)}} = \vec B_x^{(0)} + \vec B_y^{(0)} + \vec B_z^{(0)}. \\
 \end{array}
\end{equation}
\mbox{}\\
Such EM fields satisfy the Helmholtz equation. If the circular
polarized HFGW propagates along the z-direction, then the
non-vanishing perturbative EM fields are $\vec {E}_x^{(1)} $,$\vec
{B}_y^{(1)} $(the perturbative EM fields produced by the $\oplus $
polarization component of the HFGW) and $\vec {E}_y^{(1)} $,$\vec
{B}_x^{(1)} $(the perturbative EM fields generated by the $\otimes $
polarization component of the HFGW) in our scheme[32], respectively, i.e.,\mbox{}\\
\begin{equation}
\label{eq36}
\begin{array}{l}
 {{\vec E}^{(1)}} = \vec E_x^{(1)} + \vec E_y^{(1)}, \\
 {{\vec B}^{(1)}} = \vec B_x^{(1)} + \vec B_y^{(1)}. \\
 \end{array}
\end{equation}
\mbox{}\\
In this case, Eq.(25) has following concrete expression\mbox{}\\
\begin{equation}
\label{eq37}
\begin{array}{l}
 {{\vec n}_\gamma } = \frac{1}{{{\mu _0}\hbar {\omega _e}}}\vec E \times \vec B \\
  = \frac{1}{{{\mu _0}\hbar {\omega _e}}}\left\{ {\left( {\vec E_x^{(0)} + \vec E_x^{(1)} + \vec E_y^{(0)} + \vec E_y^{(1)}} \right) \times \left( {\vec B_x^{(0)} + \vec B_x^{(1)} + \vec B_y^{(0)} + \vec B_y^{(1)} + \vec B_z^{(0)}} \right)} \right\}. \\
 \end{array}
\end{equation}
\mbox{}\\
From Eq. (38), under the resonant state ($\omega_e=\omega_g$) the
total photon flux densities in the z-direction (the longitudinal
direction of the GB) and in the transverse direction (the x- and y-
directions) can be given by\mbox{}\\
\begin{equation}
\label{eq38}
\begin{array}{l}
 {n_z} = \frac{1}{{2{\mu _0}\hbar {\omega _e}}}{\mathop{\rm Re}\nolimits} \left\{ {\left[ {E_x^{*(0)}B_y^{(0)} + E_y^{*(0)}B_x^{(0)}} \right]} \right. \\
 {\rm{     ~~~                 }} + \left[ {E_x^{*(0)}B_y^{(1)} + E_y^{*(0)}B_x^{(1)} + E_x^{*(1)}B_y^{(0)} + E_y^{*(1)}B_x^{(0)}} \right] \\
 {\rm{   ~~~              }} + \left. {\left[ {E_x^{*(1)}B_y^{(1)} + E_y^{*(1)}B_x^{(1)}} \right]} \right\} \\
  \\
 {\rm{ ~~~  }} = n_z^{(0)} + n_z^{(1)} + n_z^{(2)} \\
 {\rm{  ~~~= }}n_z^{(0)} + n_z^{(1)} + o({h^2}), \\
  \\
 \end{array}
\end{equation}
\begin{equation}
\mbox{}\\
\label{eq39}
\begin{array}{l}
 {n_x} = \frac{1}{{2{\mu _0}\hbar {\omega _e}}}{\mathop{\rm Re}\nolimits} \left[ {E_y^{*(0)}B_z^{(0)} + E_y^{*(1)}B_z^{(0)}} \right] = n_x^{(0)} + n_x^{(1)}, \\
 {\rm{                          }} \\
 \end{array}
\end{equation}
\begin{equation}
\label{eq40}
\begin{array}{l}
 {n_y} = \frac{1}{{2{\mu _0}\hbar {\omega _e}}}{\mathop{\rm Re}\nolimits} \left[ {E_x^{*(0)}B_z^{(0)} + E_x^{*(1)}B_z^{(0)}} \right] = n_y^{(0)} + n_y^{(1)}. \\
 {\rm{                          }} \\
 \end{array}
\end{equation}\\

(1)The photon flux in the z-direction (the longitudinal direction
\mbox{}\\~\mbox{}~~~~~~~\mbox{}~ of
the GB)\\

From Eq. (\ref{eq38}) and Refs.[32,40], we have
\begin{equation}
\label{eq41}
n_z^{(0)} = {\left| {n_z^{(0)}} \right|_{\max }}\exp ( -
\frac{{2{r^2}}}{{{W^2}}}),{\rm{   ~~}}n_z^{(1)} = {\left|
{n_z^{(1)}} \right|_{\max }}\exp ( - \frac{{{r^2}}}{{{W^2}}}),
\end{equation}
where r is the radial distance to the symmetrical axis (the z-axis)
of the GB, $W$ is the spot radius of the GB. Eq.(\ref{eq41}) shows
that $n_z^{(0)} $ decays by the typical Gaussian decay rate $\exp
(-\frac{2r^2}{W^2})$, while $n_z^{(1)} $ decays by the factor $\exp
(-\frac{r^2}{W^2})$, i.e., the decay rate of $n_z^{(1)} $ is slower
than that of $n_z^{(0)} $. However, since $\left| {n_z^{(0)} }
\right|_{\max } \gg \left| {n_z^{(1)} } \right|_{\max } $ in almost
all of the regions (see Fig.5), it is difficult to generate an
observable effect by $n_z^{(1)} $ in these regions. For the HFGW
parameters of $h=10^{-30}$, $\nu =5\mbox{GHz}$, only if $r\to
34\mbox{cm}$ (at the xy plane), $n_z^{(1)} $ has comparable order of
magnitude with $n_z^{(0)} $. However, $n_z^{(1)} $ and $n_z^{(0)} $
all are decayed to the very small undetectable value $n_z^{(1)} \sim
n_z^{(0)} \sim 10^{-16}\mbox{s}^{-1}\mbox{m}^{\mbox{-2}}$.

\newpage

\begin{figure}[htbp]
\centerline{\includegraphics[bb=0 0 305 271]{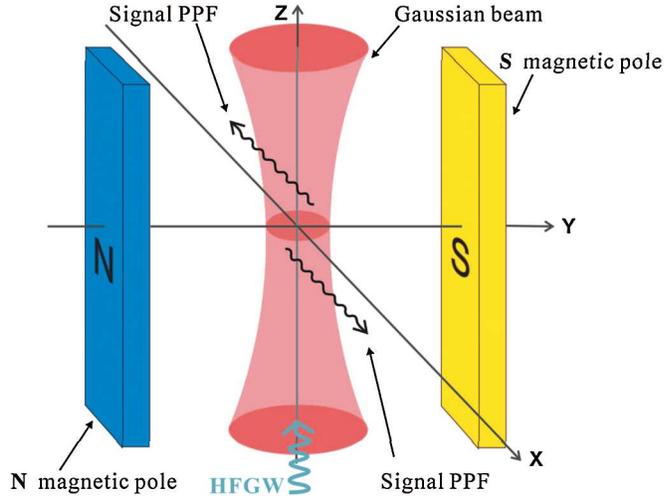} }\label{fig4}
\caption{ When the HFGW propagates along the z-direction in the
coupling system of the GB and the transverse static magnetic field
$\hat {\vec {B}}_y^{(0)} ${\small , the resonant interaction
(}$\omega _e =\omega _g ${\small ) of the HFGW with the EM fields
will generate not only the longitudinal perturbative photon flux
$n_z^{(1)}$, but also the transverse perturbative photon fluxes
(}$n_x^{(1)} ${\small and }$n_y^{(1)} ${\small ) in the x- and y-
directions due to the spread property of the GB itself. This is an
important difference between Fig.2 and Fig.4. Moreover, unlike
}$n_z^{(1)} ${\small and }$n_z^{(0)} ${\small , }$n_x^{(1)} ${\small
and }$n_x^{(0)} ${\small have very different distribution and the
decay rates.}}
\end{figure}
\newpage

\begin{figure}[htbp]
\leftline{\includegraphics[scale=0.6]{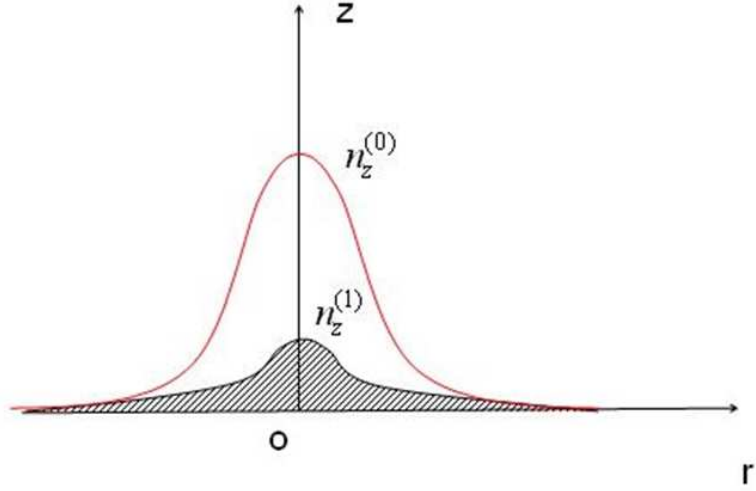}} \label{fig5}
\caption{The first-order PPF density $n_z^{(1)} ${\small and the BPF
density $n_z^{(0)} ${\small have the same propagating direction and
the similar distribution. Thus }$n_z^{(0)} ${\small is much larger
than }$n_z^{(1)} ${\small in most of the regions.}}}
\end{figure}

\mbox{}\\
 (2)The photon fluxes
in the x-direction (the transverse direction of the GB).\mbox{}\\

According to Eq. (\ref{eq39}) and Refs.[32,40], one finds
\begin{equation}
\label{eq42}
{n_x} = \frac{1}{{2{\mu _0}\hbar {\omega _e}}}{\left\langle {|\vec
E_y^{(0)}||\vec B_z^{(0)}| + |\vec E_y^{(1)}||\vec B_z^{(0)}|\cos
\delta } \right\rangle _{{\omega _e} = {\omega _g}}},
\end{equation}
Setting $\delta=0$ will always be possible by regulating the phase
of the GB. Then
\begin{equation}
\label{eq43}
\begin{array}{l}
 {n_x} = \frac{1}{{2{\mu _0}\hbar {\omega _e}}}{\left\{ {\left\langle {\vec E_y^{(0)}\vec B_z^{(0)}} \right\rangle  + \left\langle {\vec E_y^{(1)}\vec B_z^{(0)}} \right\rangle } \right\}_{{\omega _e} = {\omega _g}}} \\
  \\
  = n_x^{(0)} + n_x^{(1)} = {{\dot N}_{0x}} + {{\dot N}_{1x}} \\
 \end{array}
\end{equation}
where
\begin{equation}
\label{eq44}
n_x^{(0)} = {\dot N_{0x}} = \frac{1}{{2{\mu _0}\hbar {\omega
_e}}}\left\langle {|\vec E_y^{(0)}||\vec B_z^{(0)|}} \right\rangle
= |n_x^{(0)}{|_{\max }}x\exp ( - \frac{{2{x^2}}}{{{W^2}}}),
\end{equation}
\begin{equation}
\label{eq45}
n_x^{(1)} = {\dot N_{1x}} = \frac{1}{{2{\mu _0}\hbar {\omega
_e}}}{\left\langle {|\vec E_y^{(1)}||\vec B_z^{(0)}}| \right\rangle
_{{\omega _e} = {\omega _g}}} = {\left| {n_x^{(1)}} \right|_{\max
}}\exp ( - \frac{{{x^2}}}{{{W^2}}}),
\end{equation}
Unlike the case of plane EMW, Eqs. (45) and (46) show that $\dot
{N}_{0x} $ will be not always larger than $\dot {N}_{1x} $. In the
case of GB, $B_z^{(0)} $ of the GB depends not only on $\vec
{E}_y^{(0)} $, but also $\vec {E}_x^{(0)} $, i.e.,
\begin{equation}
\label{eq46}
B_z^{(0)} = \frac{i}{{{\omega _e}}}(\frac{{\partial
E_x^{(0)}}}{{\partial y}} - \frac{{\partial E_y^{(0)}}}{{\partial
x}}).
\end{equation}
Therefore, when $
E_y^{(0)} = 0$,$
n_x^{(0)} $ must be vanish, but $
n_x^{(1)} = n_{x\max }^{(1)} \ne 0$

\begin{figure}[htbp]
\centerline{\includegraphics[scale=0.7]{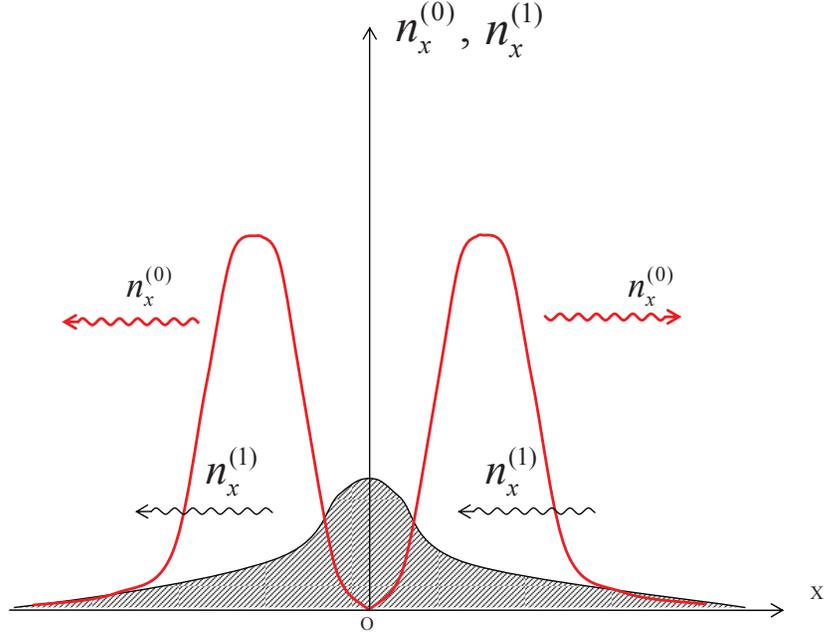}} \label{fig6}
\caption {Schematic diagram of strength distribution of  $n_x^{(0)}
${\small and $n_x^{(1)} ${\small in the ``outgoing wave'' region of
the GB (another one is the ``imploding wave'' region. For an optimum
GB, such properties of the transverse BPFs in such two regions would
be ``anti-symmetric"). Unlike Fig.5, here }$n_x^{(0)} \vert _{x=0}
=0${ while }$n_x^{(1)} \vert _{x=0} =n_x^{(1)} \vert _{\max }
${\small . Therefore, }$n_x^{(1)} \Delta t${ can be effectively
larger than the background noise photon flux fluctuation
}$(n_x^{(0)} \Delta t)^{1/2}${\small , i.e., }$n_x^{(1)} \Delta
t>(n_x^{(0)} \Delta t)^{1/2}${\small at the yz-plane and at the
parallel surfaces near the yz-plane, and }$n_x^{(1)}${ will be major
fraction of the total transverse photon flux passing through the
yz-plane, provided thermal photon flux and other noise photon fluxes
passing through the surface can be effectively suppressed. Clearly,
the EM response of the coupling system between the plane EMW and the
static magnetic field has no such characteristic. Moreover, the
propagating directions of $n_x^{(1)}$ are opposite in the regions of
$y>0$ and $y<0$ for our scheme. Thus, the total momentum of the PPF
in the x-direction vanishes. In other words, such a property ensured
conservation of the total momentum in the coherent resonance
interaction (see Ref.[32]).}}}
\end{figure}

Although Eqs. (\ref{eq44}) and (\ref{eq45}) all represent the
transverse photon fluxes in the x-direction, but their physical
behaviors are quite different:

\begin{enumerate}
\item At the yz-plane $n_x^{(1)} \vert _{x=0} =n_x^{(1)} \vert _{\max } $ where $n_x^{(0)} \vert _{x=0} =0$, i.e., the transverse PPF has a maximum at the longitudinal symmetrical surface of the GB where the transverse BPF vanishes. It should be pointed out that, the transverse BPF at the longitudinal symmetrical surfaces being identically to zero is a fundamental characteristics of the GB's, whether the circular or elliptic GB's. Thus the transverse PPF would be a major fraction of the total transverse photon fluxes flux passing through such a surface, provided the other noise photon flux passing through the surface can be effectively suppressed, although the PPF is much less than the BPF in other regions, and the PPF is always accompanied simultaneously by the BPF.
\item The $n_x^{(1)} $ and $n_x^{(0)} $ have different decay rates in the x-direction, i.e., $n_x^{(1)} \propto \exp (-\frac{x^2}{\mbox{w}^2}),n_x^{(0)} \propto x\exp (-\frac{2x^2}{\mbox{w}^2})$. The position of a maximum of $n_x^{(1)} $ is the yz plane (x=0), while the position of maximum of $n_x^{(0)} $ is about x=3.2cm in our case. Thus, SNR $n_x^{(1)} $/$n_x^{(0)} $ will be very different at the different receiving surfaces.
This means that it is always possible to obtain a best SNR
$n_x^{(1)} $/$n_x^{(0)} $ by choosing the suitable region and the
receiving surface. Using Eqs.(45) and (46), the total transverse
photon fluxes passing through the receiving surface $\Delta s$ can
be given by
\end{enumerate}
\begin{equation}
\label{eq47}
N_x^{(1)} = \int\limits_{\Delta s} {n_x^{(1)}ds,}
\end{equation}

\begin{equation}
\label{eq48}
N_x^{(0)} = \int\limits_{\Delta s} {n_x^{(0)}ds,}
\end{equation}
\mbox{}\\
 In the current scheme, $\Delta s \approx {10^{ -
2}}{{\rm{m}}^{\rm{2}}}$.
\mbox{}\\

\noindent4-3 Numerical estimation of the transverse photon fluxes.\mbox{}\\

In order to measure $N_x^{(1)} $ at a suitable receiving surface,
$N_x^{(1)} \Delta t$ (notice that here $N_x^{(1)} $ is equivalent to
$2(\dot {N}_0 \dot {N}_{GW} )^{\frac{1}{2}}$ in the plane EMW case,
but $N_x^{(1)} $ in our case and $2\left( {\dot {N}_0 \dot {N}_{gw}
} \right)^{1/2}$ in the plane EM case have a very different physical
behavior) must be effectively larger than the noise photon
fluctuation
$(N_x^{(0)} \Delta t)^{1/2}$, i.e.,\mbox{}\\
\begin{equation}
\label{eq49}
N_x^{(1)}\Delta t > {(N_x^{(0)}\Delta t)^{1/2}},
\end{equation}
then
\begin{equation}
\label{eq50}
\Delta t > {{N_x^{(0)}} \mathord{\left/
 {\vphantom {{N_x^{(0)}} {{{(N_x^{(1)})}^2}}}} \right.
 \kern-\nulldelimiterspace} {{{(N_x^{(1)})}^2}}} = \Delta {t_{\min }},
\end{equation}
\mbox{}\\
where $\Delta t_{\min } $ is requisite minimal signal accumulation
time at the noise background $N_x^{(0)} $. In fact, Eqs. (50) and
(51) are the exact forms from the general relation Eqs. (25), (26),
while Eq.(23) is the exact form from the general relation Eqs. (25)
(26) in the plane EMW case. In the following we list the$N_x^{(1)}
$,$N_x^{(0)} , \Delta t_{\min } $ and measurable HFGW strength
$h_{rms}$ at the different receiving surfaces. If x=0 (the
yz-plane), then $N_x^{(0)} $=0, it would be best measuring in the
region for $N_x^{(1)}.$ Of course, this does not mean that there are
no other noise photon fluxes passing through the receiving surface
$\Delta s$. In fact, scattering, diffraction and drift of the BPF
and the thermal noise caused by the BPF all can generate smaller the
noise photon fluxes passing through the surface $\Delta s$. Since
they are all caused by the BPF, they should have the same decay
factor $\exp (-\frac{2x^2}{W^2})$ as the BPF. Moreover, external EM
noise and the thermal noise caused by the environmental temperature
are independent of the BPF, but they can be effectively suppressed
by high-quality Faraday cage or shielding covers and low-temperature
(T$\sim $ 1K or less) vacuum operation. In general, they are much
less than the BPF. Issues such as the thermal noise, the radiation
press noise, and the noise caused by the scattering for this scheme
have been discussed in Ref.[41], we sall not repeat them here. Thus,
our attention will be focused only on the BPF itself and the other
noise photon flux $N_{x(other)}^{(0)} $ caused by the BPF. In this
case, if such noise photon fluxes passing through the receiving
surface $\Delta s$ at the yz-plane can be limited a realizable
level, then we can estimate the minimal signal accumulation time
$\Delta t_{\min } $ in the noise background.

From the above discussion, Eqs.(\ref{eq47}),(\ref{eq48}) and
Ref.[32,40], the signal photon flux $N_x^{(1)}$ and the background
photon flux $N_x^{(0)}$ passing through $\Delta {s}$ are
\begin{equation}
\label{eq51}
N_x^{(1)} = {\left| {N_x^{(1)}} \right|_{\max }}\exp ( -
\frac{{{x^2}}}{{{W^2}}}),
\end{equation}
\begin{equation}
\label{eq52}
N_x^{(0)} = {\left| {N_x^{(0)}} \right|_{\max }}x\exp ( -
\frac{{2{x^2}}}{{{W^2}}}),
\end{equation}
and
\begin{equation}
\label{eq53}
N_{x(other)}^{(0)} = {\left| {N_{x(other)}^{(0)}} \right|_{\max
}}\exp ( - \frac{{2{x^2}}}{{{W^2}}}),
\end{equation}
Displaying condition in the receiving surfaces will be
\begin{equation}
\label{eq54}
N_x^{(1)}{\left( {\Delta t} \right)^{\frac{1}{2}}} \ge {\left[
{N_x^{(0)} + N_{x{\rm{(other)}}}^{(0)}} \right]^{\frac{1}{2}}},
\end{equation}
thus
\begin{equation}
\label{eq55}
\Delta t \ge \frac{{x{{\left| {N_x^{(0)}} \right|}_{\max }} +
|N_{x{\rm{(other)}}}^{(0)}{|_{\max }}}}{{|N_x^{(1)}|_{\max
}^2}}{\rm{ and ~  }}\Delta {t_{\min }} = \frac{{x{{\left|
{N_x^{(0)}} \right|}_{\max }} + |N_{x{\rm{(other)}}}^{(0)}{|_{\max
}}}}{{|N_x^{(1)}|_{\max }^2}},
\end{equation}
\mbox{}\\
where $\left| {N_x^{(0)} } \right|_{\max } \approx 1.2\times
10^{22}s^{-1}$ in the typical parameters condition of the scheme.

Considering a possible laboratory condition, we choice the typical
parameters in Ref.[32], i.e., $\hat{B}_y^{(0)}=3T, L=6m, P=10W$.
Then we can estimate $\Delta t_{\min } $ in the different HFGW
parameters conditions.
\mbox{}\\

(1) x=0, then $
N_x^{(0)} \equiv 0 $, from Eqs. (53) and (\ref{eq55})
\begin{equation}
\label{eq56}
\Delta {t_{\min }} = \frac{{|N_{x{\rm{(other)}}}^{(0)}{|_{\max
}}}}{{|N_x^{(1)}|_{\max }^2}}.
\end{equation}
\mbox{}\\

\begin{equation}
\label{eq57}
\begin{array}{l}
If ~
 \hat h = {10^{ - 30}},{\rm{ then    }}N_x^{(1)} = |N_x^{(1)}{|_{\max }} \approx 8.2 \times {10^2}{{\rm{s}}^{ - 1}}{\rm{   and}} \\
 {\rm{                }}\Delta {t_{\min }} \approx 3.0 \times {10^3}s{\rm{   ~ provided ~  }}|N_{x{\rm{(other)}}}^{(0)} |_{max}< 2.1 \times {10^{9}}{{\rm{s}}^{{\rm{ - 1}}}}, \\
 {\rm{                }}\Delta {t_{\min }} \approx 3.0 \times {10^5}s\thicksim 3.5{\rm{~days ~ provided ~   }}|N_{x{\rm{(other)}}}^{(0)} |_{max} < 2.1 \times {10^{11}}{{\rm{s}}^{{\rm{ - 1}}}}. \\
 {\rm{                                                                                         }}~~~~~~~~~~~~~~~~~~~~~~~~~~~~~~~~~~~~~~~~~~~~~~~~~~~~~~~~~~~~~~~~~~~~~~~~~~~~~~~~(\thicksim 0.7{\rm{PW}}) \\
 \end{array}
\end{equation}

\begin{equation}
\label{eq59}
\begin{array}{l}
 \hat h = {10^{ - 27}}, {\rm{ then    }}|N_x^{(1)}{|_{\max }} \approx 8.2 \times {10^5}{{\rm{s}}^{ - 1}}{\rm{   and}} \\
 {\rm{                }}\Delta {t_{\min }} \approx 3.0 \times {10^3}s{\rm{  ~  provided ~ }}|N_{x{\rm{(other)}}}^{(0)} |_{max}< 2.1 \times {10^{15}}{{\rm{s}}^{{\rm{ - 1}}}}, \\
 {\rm{                }}\Delta {t_{\min }} \approx 3.0 \times {10^5}s\thicksim 3.5{\rm{~days~  provided~    }}|N_{x{\rm{(other)}}}^{(0)} |_{max} < 2.1 \times {10^{17}}{{\rm{s}}^{{\rm{ - 1}}}}. \\
 \end{array}
\end{equation}

\begin{equation}
\label{eq58}
\begin{array}{l}
 \hat h = {10^{ - 26}},{\rm{ then    }}|N_x^{(1)}{|_{\max }} \approx 8.2 \times {10^6}{{\rm{s}}^{ - 1}}{\rm{   and}} \\
 {\rm{                }}\Delta {t_{\min }} \approx 3.0 \times {10^3}s{\rm{ ~   provided ~ }}|N^{(0)}_{x{\rm{(other)}}}|_{max} < 2.1 \times {10^{17}}{{\rm{s}}^{{\rm{ - 1}}}}, \\
 {\rm{                }}\Delta {t_{\min }} \approx 3.0 \times {10^5}s\thicksim 3.5{\rm{~days  ~provided  ~  }}N_{x{\rm{(other)}}}^{(0)} < 2.1 \times {10^{19}}{{\rm{s}}^{{\rm{ - 1}}}}. \\
 \end{array}
\end{equation}

\begin{equation}
\label{eq60}
\begin{array}{l}
 \hat h = {10^{ - 24}},{\rm{ then    }}|N_x^{(1)}{|_{\max }} \approx 8.2 \times {10^8}{{\rm{s}}^{ - 1}}{\rm{   and}} \\
 {\rm{                }}\Delta {t_{\min }} \approx 3.0 \times {10^3}s{\rm{   ~ provided~  }}|N_{x{\rm{(other)}}}^{(0)}|_{max} < 2.1 \times {10^{21}}{{\rm{s}}^{{\rm{ - 1}}}}, \\
 {\rm{                }}\Delta {t_{\min }} \approx 3.0 \times {10^5}s\thicksim3.5{\rm{~days~  provided ~   }}|N_{x{\rm{(other)}}}^{(0)}|_{max} < 2.1 \times {10^{23}}{{\rm{s}}^{{\rm{ - 1}}}}. \\
 \end{array}
\end{equation}

The above results show that limitation to the other noise photon
fluxes passing through $\Delta s$ would be very relaxed. It is
interesting to compare the scheme employed earlier (see Eq. (23),
where $h=10^{-26}, \nu=5GHz, \hat{B}^{(0)}_y=10T, L=10m, P=10W$) and
the current scheme (see Eq.(60), here $\hat{h}=10^{-26}, \nu=5GHz,
\hat{B}^{(0)}=3T, L=6m, P=10W$), they show that the current scheme
has obvious advantages and reality.
\mbox{}\\

(2) x=1cm=10$^{-2}$m, then $N_x^{(0)} \approx 1.1\times
10^{20}\mbox{s}^{\mbox{-1}}$, but where $\vert N_{x_{(other)}}^{(0)}
\vert _{\max } $ is often much less than $N_x^{(0)} $ i.e.,
$N_{x_{(other)}}^{(0)} $ can be neglected in the all following
discussions. From Eq.(56), we have
\begin{equation}
\label{eq61}
\begin{array}{l}
 \hat h = {10^{ - 26}},{\rm{   ~~~ }}N_x^{(1)} \approx 7.8 \times {10^6}{s^{ - 1}},{\rm{  ~~~  }}\Delta {t_{\min }} \approx 1.8 \times {10^6}s. \\
 \hat h = {10^{ - 25}},{\rm{    ~~~}}N_x^{(1)} \approx 7.8 \times {10^7}{s^{ - 1}},{\rm{   ~~~ }}\Delta {t_{\min }} \approx 1.8 \times {10^4}s. \\
 \hat h = {10^{ - 24}},{\rm{    ~~~}}N_x^{(1)} \approx 7.8\times {10^8}{s^{ - 1}},{\rm{    ~~~}}\Delta {t_{\min }} \approx 1.8\times {10^2}s. \\
 \end{array}
\end{equation}
(3) $
x = 2cm = 2 \times {10^{ - 2}}m$, then $
N_x^{(0)} \approx 1.7 \times {10^{20}}{{\rm{s}}^{{\rm{ - 1}}}}$
\begin{equation}
\label{eq62}
\begin{array}{l}
 \hat h = {10^{ - 26}},{\rm{    ~~~}}N_x^{(1)} \approx 7.0 \times {10^6}{{\rm{s}}^{{\rm{ - 1}}}},{\rm{   ~~~ }}\Delta {t_{\min }} \approx 3.5 \times {10^6}s. \\
 \hat h = {10^{ - 25}},{\rm{   ~~~ }}N_x^{(1)} \approx 7.0 \times {10^7}{{\rm{s}}^{{\rm{ - 1}}}},{\rm{   ~~~ }}\Delta {t_{\min }} \approx 3.5 \times {10^4}s. \\
 \hat h = {10^{ - 24}},{\rm{   ~~~ }}N_x^{(1)} \approx 7.0 \times {10^8}{{\rm{s}}^{{\rm{ - 1}}}},{\rm{    ~~~}}\Delta {t_{\min }} \approx 3.5 \times {10^2}s. \\
 \end{array}
\end{equation}
(4)$
x = 3cm = 3 \times {10^{ - 2}}m$, then $
N_x^{(0)} \approx 1.8 \times {10^{20}}{{\rm{s}}^{{\rm{ - 1}}}}$
\begin{equation}
\label{eq63}
\begin{array}{l}
 \hat h = {10^{ - 26}},{\rm{ ~~~   }}N_x^{(1)} \approx 5.8 \times {10^6}{{\rm{s}}^{{\rm{ - 1}}}},{\rm{   ~~~ }}\Delta {t_{\min }} \approx 5.4 \times {10^6}s. \\
 \hat h = {10^{ - 25}},{\rm{   ~~~ }}N_x^{(1)} \approx 5.8 \times {10^7}{{\rm{s}}^{{\rm{ - 1}}}},{\rm{    ~~~}}\Delta {t_{\min }} \approx 5.4 \times {10^4}s. \\
 \hat h = {10^{ - 24}},{\rm{    ~~~}}N_x^{(1)} \approx 5.8 \times {10^8}{{\rm{s}}^{{\rm{ - 1}}}},{\rm{    ~~~}}\Delta {t_{\min }} \approx 5.4 \times {10^2}s. \\
 \end{array}
\end{equation}

\noindent(5)$
x = 10cm = 0.1m$, $
N_x^{(0)} \approx 4.0 \times {10^{17}}{{\rm{s}}^{{\rm{ - 1}}}}$
\begin{equation}
\label{eq65}
\begin{array}{l}
 \hat h = {10^{ - 26}},{\rm{~~~    }}N_x^{(1)} \approx 1.5 \times {10^5}{s^{ - 1}},{\rm{~~~     }}\Delta {t_{\min }} \approx 1.2 \times {10^7}s. \\
 \hat h = {10^{ - 25}},{\rm{   ~~~  }}N_x^{(1)} \approx 1.5 \times {10^6}{s^{ - 1}},{\rm{  ~~~   }}\Delta {t_{\min }} \approx 1.2 \times {10^5}s. \\
 \hat h = {10^{ - 24}},{\rm{  ~~~   }}N_x^{(1)} \approx 1.5 \times {10^7}{s^{ - 1}},{\rm{  ~~~   }}\Delta {t_{\min }} \approx 1.2 \times {10^3}s. \\
 \end{array}
\end{equation}

\noindent(6) $x=29\mbox{cm}$, (about distance of 6 spot radiuses of
the GB), \mbox{}\\ \mbox{}  ~  ~  $\hat {h}=10^{-26}$, then
$N_x^{(0)} \approx N_x^{(1)} \approx 2.1\times
10^{-8}\mbox{s}^{\mbox{-1}}$. Time of \mbox{}\\ \mbox{} ~~~receiving
one transversal photon would be $\Delta t_{\min } \approx
\frac{1}{N_x^{(0)} }\approx \frac{1}{N_x^{(1)}
}\mbox{}\\~\mbox{}~~\approx \frac{1}{2.1\times
10^{-8}\mbox{s}^{\mbox{-1}}}=4.8\times 10^7$ s.\mbox{}\\

The above numerical estimation shows that:

\begin{enumerate}
\item The best position for displaying $N_x^{(1)} $ would be in the yz-plane and the other parallel receiving surfaces in the region of $-2cm<x<2cm$. In such regions, the transverse PPF $N_x^{(1)} $ for the parameter condition $\hat {h}\sim 10^{-24}-10^{-30}$ may reach up to $\sim 8.2\times 10^8s^{-1}$ to $8.2\times 10^2s^{-1}$. If other noise photon fluxes passing through the surfaces can be effectively suppressed into $\sim 2.1\times 10^{23}s^{-1}$ to $\sim 2.1\times 10^{9}s^{-1},$ then corresponding minimal signal accumulation
time $\Delta t_{\min } $ in the noise photon flux background would
be $\sim10^3 s ~$to $10^5s$.
\item Unlike $N_x^{(1)} $,$N_x^{(0)} $ has maximum at $x\sim 3.2cm$, where $N_x^{(0)} \gg N_x^{(1)} $, but $N_x^{(1)} \vert _{x=3.2cm}  \mbox{ and }  N_x^{(1)} \vert _{x=0} =\mbox{ }N_x^{(1)} \vert _{\max } $ have the same order of magnitude. In the region, the detecting sensitivity would be worse by 3-4 orders of magnitude over that at the yz-plane.

\item Since $N_x^{(1)} = {\rm{ |}}N_x^{(1)}{|_{\max }}\exp ( - \frac{{{x^2}}}{{{W^2}}})$,$N_x^{(0)} = {\rm{ |}}N_x^{(0)}{|_{\max }}x\exp ( - \frac{{2{x^2}}}{{{W^2}}})$, even if $\hat {h}$=10$^{-26}$, they will have the same order of magnitude in $x\approx 29cm$. However, where $N_x^{(0)} ,N_x^{(1)} $ all decay
to $2.1\times 10^{-8}\mbox{s}^{\mbox{-1}}$.
\end{enumerate}
\mbox{}\\

Moreover, it was shown that if the propagating detections of
$N_x^{(0)}$ and $N_x^{(1)}$ are the same in 1st, 3rd, 6th and 8th
octants in our case, then they will propagate along the opposite
directions in the 2nd, 4th, 5the and 7the octants [32]. This means
the distinguishing ability to $N_x^{(0)}$ and $N_x^{(1)}$ of the
scheme can be further improved. Also, as suggested by Baker [42],
since the BPF is unaffected by the magnetic field (it is only
involved in the generation of the PPF), one can differentiate the
PPF from the BPF by modulating the magnetic field. This essentially
eliminates the BPF by microwave-receiver signal processing. For
example, one measures the BPF plus PPF with the magnet on and then
measures the BPF alone with the magnet off and subtracts one from
the other in order to obtain the PPF alone. This process is
accomplished more rigorously by statistical signal processing.\\
\\
\\

\noindent\Large{4.4 Role of fractal membranes or other equivalent

microwave lenses}\\\mbox{}
 \large{\mbox{}}\\
(1). The FMs is merely one of many possible ways to improve the SNR
and detecting quality via the redirection of signal photons onto the microwave detectors [32]. However, in the above discussion,
the proposal scheme did not involve the FMs. In order words, even if
we do not use the FMs, the above-mentioned relation between the PPF and
the BPF is still valid. The fractal membranes in the GHz band have
successfully been developed by the Hong Kong University of Science
and Technology [43-45] from 2002-2005. Firstly, the fractal
membranes (FMs) have very good selection ability to the photon
fluxes in the GHz band. If the FM is nearly totally reflecting for
the photon fluxes with certain frequencies in the GHz band, then it
will be nearly total transmitting for the photon fluxes with other
frequencies in the GHz band. Secondly, the FMs have good focus
function to the photon fluxes in the GHz band. For example, the
photon fluxes reflected and transmitted by the FMs can keep their
strength invariant within the distance of 1 meter from the FMs. Such
function has been proven by experimental tests. The role of the FMs
in the scheme is only the reflector or the transmitter for the
photon flux in the GHz band. Because $N_z^{(0)} ,N_y^{(0)} $ and
$N_x^{(0)} ,N_x^{(1)} $ are exactly orthogonal for each other, an FM
(or an equivalent microwave lens) paralle with the yz-plane
would focuses only $N_x^{(0)} ,N_x^{(1)} $ and not $N_z^{(0)}
,N_y^{(0)} $. In fact, here requirement for the FMs is also more
relaxed, i.e., it does not require focusing the photon flux onto a
micron-sized detector even into a point. In the typical parameter
condition of the scheme, if the cross section of the focusing photon
flux and the image size has the same or close size in the detector (
in distance of $\sim $28cm,) then the SNR $N_x^{(1)} /N_x^{(0)} $ at
the receiving surface $\Delta s$ and at the image surface $\Delta
s'$ would be nearly the same. Moreover, because unfocused
$N_z^{(0)},N_y^{(0)}$ will be decayed to 10$^{-7}$s$^{-1}$ at
x=29cm, their influence can be neglected there.\\

(2). If the FM is just laid at the symmetrical plane (the yz-plane)
or at the parallel planes very near the yz-plane (see Figs. 7 and
8), then the wave-fronts of the photon fluxes passing through the
receiving surfaces $\Delta s$ at the planes would be the plane or
the pseudo-plane, i.e., where it is possible to obtain a better
focusing effect. The requirement for the focus in the region would
be more relaxed than other regions. This is because such focusing
quality depends only on the local interaction of the photon fluxes
at the receiving surfaces in the region of $\vert x\vert \le 2cm$.
Besides, provided the photon fluxes focused by the FM can keep a
plane or pseudo-plane wave-front, then $N_x^{(0)} ,N_x^{(1)} $
focused simultaneously on another surface $\Delta s'$ would have the
same or nearly the same SNR as with that at $\Delta s$. A unique
requirement for $N_x^{(1)} $ and $N_x^{(0)} $ at $\Delta s'$ is that
$N_x^{(1)} \left( {\Delta t} \right)^{\frac{1}{2}}$ should be larger
than $\sqrt {N_x^{(0)} } $ in a typical experimental time interval
$\Delta t$, and this process does not need an image of high-quality
at $\Delta s'$. Contrarily, if the FM is laid at an obvious nonsymmetrical plane, then it is difficult to focus the photon fluxes due to the spread property of the GB (see Fig. 9).\\

(3). The photon fluxes $N_z^{(0)} $ and $N_z^{(1)} $ in the
z-direction have a similar property. However, unlike the relation
between $N_x^{(0)} $ and $N_x^{(1)} $, $N_z^{(0)} $ (noise) is much
larger than $N_z^{(1)} $ (signal) in the almost of all regions. This
is a very important difference between the photon fluxes in such two
directions.\\

 (4). A major role of the FM or other equivalent
microwave lenses in the scheme is their focusing effect and not
their superconductivity, and this does not mean that one can measure
only $N_x^{(1)} $ (``interference term'') and not $N^{(0)}$
(background). Also, it does not mean that $N^{(0)}$ is neglected and
$N^{(0)}$ does not reach the photon flux detector. Actually, the FM
is immersed in the BPF. Thus the BPF will generate the thermal noise
in the FM. However, the BPF itself and the thermal noise photons
caused by the BPF in the FM have an essential difference. The former is
vector and has high directivity; the latter are photons of random
thermal motion. Under the low-temperature condition, the latter are
much less than the former. In particular, $N_z^{(0)} ,N_y^{(0)} $ of
the BPF are exactly parallel to the yz-plane and exactly
perpendicular to $N_x^{(0)} $ and $N_x^{(1)} $.Thus $N_z^{(0)} $ and
$N_y^{(0)} $ do not provide any direct contribution to the photon
flux passing through the receiving surfaces parallel to the yz
plane, nor are they reflected, transmitted or focused by
the FMs laying at the receiving surfaces. In other words, the photon
flux focused by the FM will be $N_x^{(0)} ,N_x^{(1)} $ and not
$N_z^{(0)} ,N_y^{(0)} $. In this case $N_x^{(1)} $ and $N_x^{(0)} $
would reach simultaneously the detector, but $N_x^{(1)} $ and
$N_x^{(0)} $ in the different receiving surfaces have the different
ratio$N_x^{(1)} /N_x^{(0)} $, this is an important difference to the
plane EMW case. Therefore, it is always possible to choose a best
region and the receiving surface to detect the total photon flux
($N_x^{(0)}
+N_x^{(1)} )$ which has a good SNR. Furthermore, the $N_x^{(0)}$ can be differentiated from the $N_x^{(1)}$ by modulating the $\hat{B}_y^{(0)}$.\\

\begin{figure}[htbp]
\centerline{\includegraphics[bb=93 550 500 744]{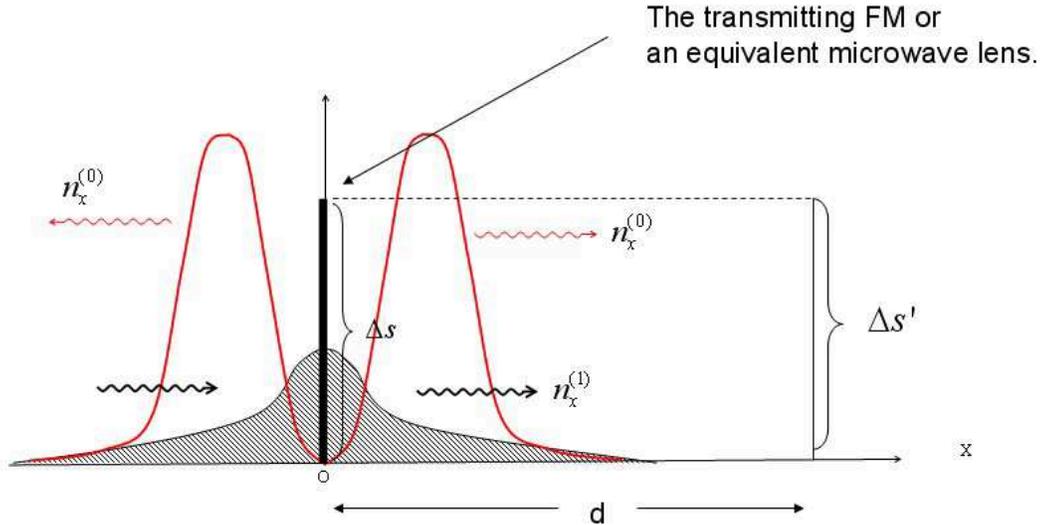}}
\label{fig7a} \caption{ Unlike the photon fluxes $N_z^{(0)}
,N_z^{(1)} ${\small ,}$N_x^{(1)} \vert _{x=0} =N_x^{(1)} \vert
_{\max } ${\small where }$N_x^{(0)} \vert _{x=0} =0${\small . This
means that }$N_x^{(0)} \mbox{ and }N_x^{(1)} ${\small focused by the
FM at the yz-plane or at the parallel planes very near the yz-plane
would have a good focusing effect and the SNR.}}
\end{figure}

\begin{figure}[htbp]
\centerline{\includegraphics[bb=90 535 500 744]{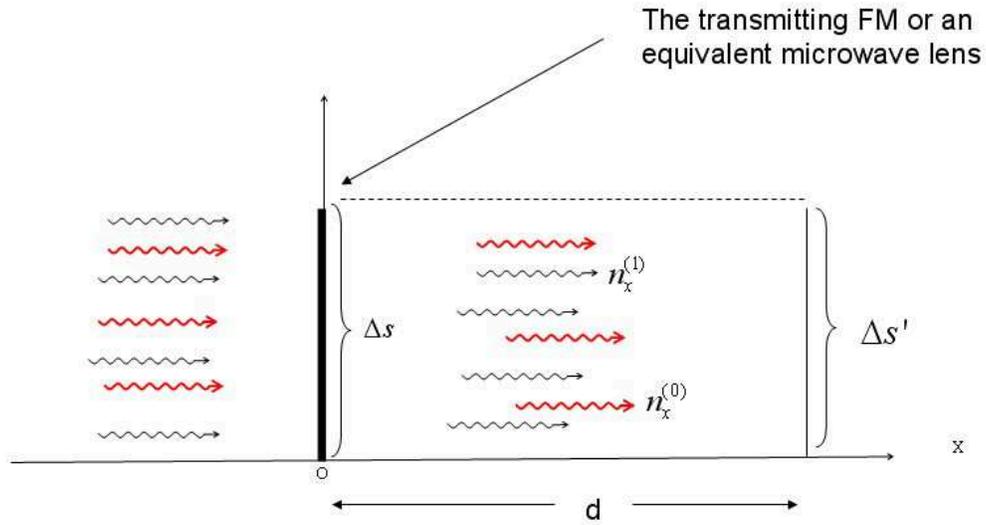}}
\label{fig7b} \caption{If the FM is just laid at the yz-plane or at
the parallel planes very near the yz-plane, then the wave-fronts of
the photon fluxes passing through the planes would be the plane or
the pseudo-plane, and it is possible to obtain an effective focusing
effect.}
\end{figure}

\begin{figure}[htbp]
\centerline{\includegraphics[bb=90 535 500 744]{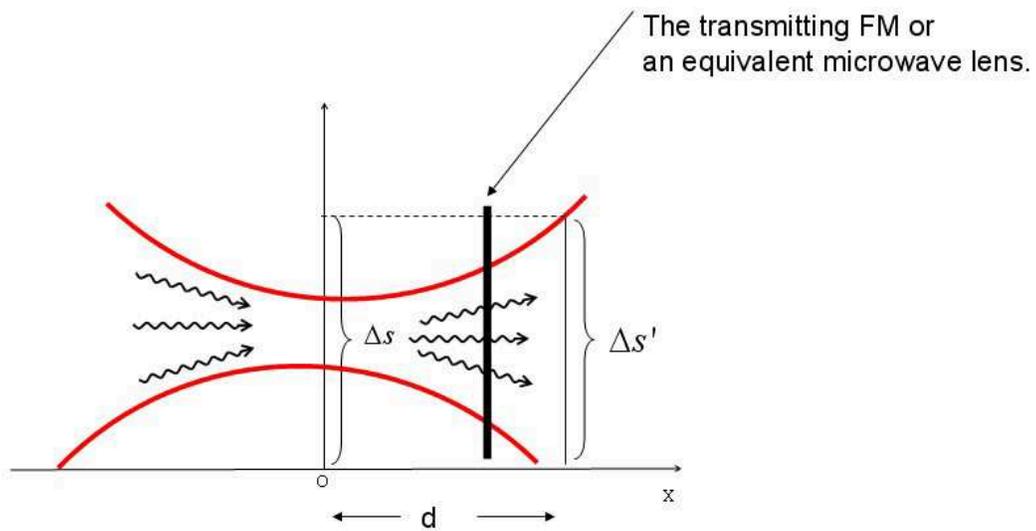}}
\label{fig7c} \caption{ If the FM is laid at an obvious
non-symmetrical plane, then it is difficult to focus the photon
fluxes due to the spread property of the GB.}
\end{figure}

\mbox{}\\
\mbox{}\\

 \noindent\Large{4.5  Challenge and issues}\\\mbox{}
 \large{\mbox{}}

Except for the above-principle analysis, of course, one must
consider following challenge and issues. They would include the
generation of high-quality GB, suppression of the noises, such as
thermal noise, the radiation press noise, noises caused by the
scattering of photons, dielectric dissipation due to the dust and
other particles, and the concrete influence and correction of the
FMs to the GB itself, etc.

The low-temperature (T$\sim $1K or less) and vacuum operation can
effectively reduce the thermal noise and dielectric dissipation.
There is room for improvement in other ways as well. They would include utilization of super-strong
static magnetic fields, matching of ultra-high sensitivity microwave
photon detectors, construction of a good ``microwave darkroom'',
coupling between the open superconducting cavities and the current
scheme (the open superconducting cavities have very large quantity
factor Q$\sim $10$^{9}$-10$^{11}$, this coupling might greatly
enhance the signal photon flux and not increase obviously the noise
power), etc. All these issues need further theoretical study and
careful experimental investigation, and they would provide new ways
and possibilities to further narrow the gap between the detection schemes and the
reality of a valid measurement.\\

\Large {\Rmnum{5}. Brief summary } \large{\mbox{}}\\

The EM detecting scheme based on the pure inverse G-effect in the
laboratory  would not be capable of detecting the HFGWs in the GHz
band, while the coupling system between the Gaussian-type microwave
photon flux, the static magnetic field and the fractal membranes (or
other equivalent microwave lenses) will be a useful candidate. The
key parameter in the current scheme is not the second-order PPF but
the transverse first-order PPF; the measurable photon flux is not
only the transverse first-order PPF but the total transverse photon
flux, and they have different SNRs at the different receiving
surfaces; the requisite minimal accumulation time $\Delta t$ of the
signal at the special receiving surfaces and in the background
photon flux noise would be $\sim $10$^{3}$-10$^{5}$ seconds for the
typical laboratory condition and the parameters of $\hat {h}\sim
10^{-26}-10^{-30}$ at $\nu =5\mbox{GHz}$ with bandwidth $\sim$1Hz

This paper does not involve the standard quantum limit (SQL) caused
by the quantum back-action. The SQL constrains the lowest possible
sensitivity. We shall show that the SQL in the current scheme does
not constrain predicated sensitivity (including the constant
amplitude HFGWs and the stochastic high-frequency relic GWs). In
other words, the sensitivity in the current scheme is the photon
signal limited, not quantum noise limited[46]. We will discuss
relative issues elsewhere.
\newpage
\begin{center}
\Large {Acknowledgements } \large{\mbox{}}\\
\end{center}
\mbox{}\\
We would like to thank Dr. A. Beckwith for his very useful
discussions and suggestion. This work is supported by the National
Nature Science Foundation of China under Grant No. 10575140, the
Foundation of China Academy of Engineering Physics under Grant No.
2008T0401, 2008T0402, Chongqing University Postgraduates Science and
Innovation Fund, Project Number. 200811B1A0100299,
GRAVWAVE{\textregistered}LLC, Transportation Science Corporation and
Seculine Consulting of the USA.
\newpage

\begin{center}
References
\end{center}

\noindent[1]R.L.Forward and R.M.L.Baker," Gravitational gradients,
gravitational \\~\mbox{}~~~waves and the 'Weber bar'", Lecture at
Lockheed Astrodynansics \\~\mbox{}~~~Research Center, Bel
Air,California,650N.Sepulveda,Bel Air,California, \\~\mbox{}~~~USA,
November 16th, Lockheed Research Report RL 15210(Forward
\\~\mbox{}~~~coined the term "High Frequency Gravitational
Waves").(1961).

\noindent[2]M.E.Gertsenshtein, Sov.Phys. JETP\textbf{64},84 (1962)

\noindent[3]L.Halpern and B.Laurent,Nuovo
Cimento\textbf{33},728(1964).

\noindent[4]R.A.Isaason,Phys.Rev.\textbf{166},1263(1968)

\noindent[5]R.A.Isaason,Phys.Rev.\textbf{166},1272(1968)

\noindent[6]L.P. Grishchuk and
M.V.Sazhin,Sov.Phys.JETP\textbf{32},213(1974)

\noindent[7]L.P. Grishchuk and
M.V.Sazhin,Sov.Phys.JETP\textbf{41},787(1975)

\noindent[8]G.F. Chapline,J. Nuckolls and
L.L.Wood,Phys.Rev.D\textbf{10},1064(1974)

\noindent[9]V.B. Braginsky and V.N. Rudenko, Phys.Reports
\textbf{46},166(1978)

\noindent[10]S.W. Hawking and W. Israd, General Relativity: An
Einstein

\mbox{}~Centenary Survoy(Cambridge: Cambridge University
Press),90-137(1979)

\noindent[11]M. Giovannini, Phys. Rev. D \textbf{60},123511(1999)

\noindent[12]M. Giovannini, Class.Quantum
Grav,\textbf{16},2905(1999)

\noindent[13]A. Riagudo and J.P. Ugan,Phys, Rev.D\textbf{62},
083506(2000)

\noindent[14]M, Giovannini, Phy.Rev.D\textbf{73}, 083305(2006)

\noindent[15]M,Giovannini,astro-ph/0807(2008)

\noindent[16]J.E.Lidsey et al.,Phys,Reports\textbf{337},343(2000)

\noindent[17]M. Gasperini and G. Veneziano, Phys. Reports
\textbf{373},1(2003)

\noindent[18]G.Veneziano,Sci.Am.\textbf{290},30(2004)

\noindent[19]G.S.B.Kogan and V.R.Rudenko, Class Quantum Grav
21,3347(2004)

\noindent[20]R.M.L.Baker,R.C.Woods and
F.Y.Li,AIP.Proc.\textbf{813},1280(2006)

\noindent[21]P.Chen,Mod.Phys.Lelt.A\textbf{6},1069(1991)

\noindent[22]A.I.Nikishov and V.I.Ritus,Sov.
Phys.JETP\textbf{69},876(1989)

\noindent[23]A.I.Nikishov and V.I.Ritus,Sov. Phys.JETP71,643(1990)

\noindent[24]G. Gratta et al.,Workshop on Beam-Beam and
Beam-Radiation

\mbox{}~Interaction; High Intensity and Nonlininear Effects,Los
Angeles,USA,1991,

\mbox{}~edited by C. Pellegrini et al.,(World
Scientific,Singapore),70(1992)

\noindent[25]P.Chen, Resonant Photon Graviton Conversion in EM
Fidas:From

\mbox{}\mbox{}~Earth to Heaven, Stanford Linear Accelerator
Center-PUB-6666.(September,1994)

\noindent[26]X.G.Wu and Z.Y.Fang,Phys.Rev.D\textbf{78},094002(2008)

\noindent[27]A.M.Cruise, Class.Quantum Grav.\textbf{17},2525(2000)

\noindent[28]A.M. Cruise and R.M.J. Ingley, Class,Quanbum
Grav.\textbf{22}.S479 (2005)

\noindent[29]R. Ballantini et al.,gr-qc/0502054(2005)

\noindent[30]R. Ballantini et al., Class.Quantum
Grav,\textbf{20},3505(2003)

\noindent[31]A. Nishigawa et al.,Phys.Rev.D\textbf{77},022002(2008)

\noindent[32]F.Y. Li, R.M.L. Baker,Z.Y. Fang, G.V .Stepheson and
Z.Y.Chen,Eur.Phys.J.C\textbf{56},407(2008)

\noindent[33] W.K.De Logi and A.R.Mickelson, Phys. Rev.
D\textbf{16}, 2915(1977)

\noindent[34] A.N. Cillis and D.D. Harari, Phys. Rev. D\textbf{54},
4757 (1996)

\noindent[35] D.Boccaletti et al., Nuovo Cim.B \textbf{70},
129(1970)

\noindent[36] L.D. Landau and E.M. Lifshitz. The Classical Theory of
Fields.

\mbox{}~(Nauka, Moscow) 1973, PP. 368-370

\noindent[37] P.Chen, Phys. Rev. Lett. \textbf{74}, 634 (1995)

\noindent[38] M.Marklund, G. Brodin and P. Dunsby, Astrophys. J.
\textbf{536},875(2000)

\noindent[39] V.De, Sabbata et al., Sov. J. Nucl. Phys. \textbf{8},
53-7 (1969)

\noindent[40] A. Yariv, Quantum Electronics 2$^{nd}$ ed, (Wiley, New
York, 1975),

\mbox{}~P.110-129

\noindent[41]J.Li, F.Y.Li and Y.H. Zhong,Chinese Physics B
\textbf{18},922(2009)

\noindent[42] R. M. L. Baker, Private Communication, June(2008)

\noindent[43] W.J.Wen et.al., Phys. Rev. Lett. \textbf{89}, 223901
(2002)

\noindent[44] L.Zhou et al., Appt. Phys. Lett. \textbf{82}, 1012
(2003)

\noindent[45] B. Hou et al., Opt. Express \textbf{13}, 9149 (2005)

\noindent[46] G. V. Stephenson, "The standard quantum limit for the
Li-Baker

\mbox{}~HFGW detector," Proceedings of the Space, Propulsion and
Energy

\mbox{}~Sciences International Forum (SPESIF), 24-27 February,
Edited by

\mbox{}~Glen Robertson. (Paper 023), American Institute of Physics

\mbox{}~Conference Proceedings, Melville, NY, Vol. \textbf{1103},
February 2009,

\mbox{}~pp. 542-547.
\end{document}